# A New Compact Expression for the Powers of Hybrid Modes of the Step-Index Optical Fiber

W. Astar

University of Maryland, Baltimore County(UMBC); Baltimore, Maryland 21250. Email: wastar1@UMBC.edu

**Abstract:** A new, compact, and exact general expression is developed for the powers of hybrid EH- and HE-modes of the standard step-index optical fiber. The expression is found with the help of the Pauli matrices. The expression is reducible to widely used approximate forms for hybrid mode powers of weakly-guided fibers.

## 1 Introduction

The step-index optical fiber, which is a cylindrical waveguide, is widely used in the fields of optical communication and sensor technology. There are basically 2 types of fibers, the multi-mode fiber (MMF) and the single-mode fiber (SMF). The MMF can generally support a variety of modes such as the TE and the TM, respectively for which longitudinal components of the electric field ($E_z$) and the magnetic field ($H_z$) are zero. The MMF is also capable of supporting hybrid modes [1, 2]. The hybrid modes are classified as HE (EH) when the $E_z$ ($H_z$) component is dominant relative to the $H_z$ ($E_z$) component [2]. Since the mode's electromagnetic (EM) field extends into the cladding in general, mode power requires a consideration of the core along with the cladding, resulting in 12 field components.

Long-haul optical fiber transmission is generally deployed using the SMF, which is designed to support a solitary $HE_{11}$-mode, over a range of optical frequencies. The ideal SMF supports the $HE_{11}$-mode in 2 polarization states of identical eigenvalues or propagation constants ($\beta$). However, fabrication imperfections and environmental perturbations lift the degeneracy of the 2 polarization states, resulting in 2 distinct propagation constants [2].

Since hybrid modes are common to both MMFs and SMFs, the goal of this report is to develop a method of arriving at a new, compact expression for the total optical power due to a hybrid mode. The derivation is carried out with the help of the Pauli matrices [3], historically used in the analysis of light polarization. For an optical fiber of a core radius $a$, and a cladding radius $b$, the a hybrid mode's power $P$ is found to be:

$$P = \varepsilon_0 \omega_0 \beta P_0 S_J \left[ \begin{array}{c} n_1^2 \sum_{p=0}^{1} \sum_{q=-1}^{1} \xi^2 e^{ip\pi} Q(q, s_1) J_{n-p+q}(u) J_{n+p+q}(u) \\ + n_2^2 \sum_{m=1}^{2} \sum_{p=0}^{1} \sum_{q=-1}^{1} \kappa_m^2 R_n^2 e^{i(m+p)\pi} Q(q, s_2) K_{n-p+q}(w\kappa_m) K_{n+p+q}(w\kappa_m) \end{array} \right]$$

where $\omega_0$ is the radial frequency of the EM field, $S_J$, the cross-sectional area of the core, and $\kappa_m = (a/b)^{1-m}$. The expression consists of the sum of the powers in the core and in the cladding of the optical fiber, respectively represented by their refractive indices $n_1$ and $n_2$.

In the weakly-guided-fiber approximation, the expression yields either the EH-mode power, or the HE-mode power, respectively dependent on whether $s$ (which is the dispersion relation) and $g$ are both $\approx +1$, or both $\approx -1$, in the coefficient $Q(q, g) = q^2(1+qs)(1+qg)$. The expression is reducible to widely used approximate forms for hybrid mode powers of weakly-guided fibers.

## 1 A General Expression for Electromagnetic Power

The cross-section of the optical waveguide is assumed to be co-incidental with the $xy$-plane, implying that the electromagnetic (EM) field propagation is in the $z$-direction. The analysis assumes that the waveguide structure is composed of linear, homogeneous, isotropic, non-magnetic, charge-free, current-free media. The longitudinal perturbations are also assumed to be sufficiently small to render the waveguide $z$-invariant. Most practical optical waveguides can be analyzed in either rectangular (Cartesian) coordinates

$$(x, y, z) : (x, y, z) \in (-\infty, \infty) \times (-\infty, \infty) \times (-\infty, \infty) \tag{1}$$

or in cylindrical coordinates,

$$(\rho, \varphi, z) : (\rho, \varphi, z) \in [0, \infty) \times [0, 2\pi) \times (-\infty, \infty) \tag{2}$$

The following discussion applies to both coordinate systems in general. The propagating power is found from the integral of the longitudinal component of the Poynting vector, over the cross-sectional area $S$ of the optical waveguide [1]

$$P = \frac{1}{2} \operatorname{Re} \int_S \mathbf{E}^* \times \mathbf{H} \cdot \mathbf{a}_z \, dS \tag{3}$$

The Re-operator is conventionally omitted for waveguide mode propagation, due to a fortuitous cancellation of the time-dependent complex exponential in the cross-product [2]. It is observable that since the outcome of the cross-product in (3) is involved in a dot-product with the longitudinal unit-vector $\mathbf{a}_z$, the EM field components must be exclusively in the transverse direction - otherwise, the dot-product vanishes. Another equivalent interpretation, is that the computation of power only requires the Jones vectors of the electric and magnetic fields. The Jones vector is conventionally used to describe the transverse profile of the electromagnetic EM field. The EM field has a general form

$$\mathbf{U} = \mathbf{U}_{T_1} + \mathbf{U}_{T_2} + \mathbf{U}_L = \mathbf{U}_T + \mathbf{U}_z \quad \Big| \quad \mathbf{U} \in \{\mathbf{E}, \mathbf{H}\} \tag{4}$$

A transverse field may be described as a column vector

$$\mathbf{U}_T \rightarrow |\mathbf{U}_T\rangle = \begin{bmatrix} |U_{T_1}\rangle \\ |U_{T_2}\rangle \end{bmatrix} \tag{5}$$

which may be thought of as the canonical form of the Jones vector, in that the (1,1)-slot is reserved for the 1st transverse component, while the other (2,1)-slot, for the 2nd transverse vector component. The $x$-component is designated as the 1st transverse component for the Cartesian coordinate system, whereas the $\rho$-component, the 1st transverse component of the cylindrical coordinate system. This convention is used throughout this report.

If one of the fields of the EM field is known, the other may be found from one of Maxwell's curl equations in the Fourier domain [2]. The EM field is assumed to have no temporal dependence beyond continuous-wave (CW) propagation. If $\mathbf{E}$ were known *a priori*, then in the Fourier domain

$$\nabla \times \mathbf{E} = -j\omega_0 \mu_0 \mathbf{H} \tag{6}$$

where $\omega_0$ is the radial frequency of the EM field. The del-operator may be decomposed into its transverse and longitudinal components [4]:

$$\nabla = \nabla_T + \nabla_z = \frac{\mathbf{a}_{T_1}}{h_1} \frac{\partial}{\partial r_{T_1}} + \frac{\mathbf{a}_{T_2}}{h_2} \frac{\partial}{\partial r_{T_2}} + \mathbf{a}_z \frac{\partial}{\partial r_z} \tag{7}$$

This expression is general, and is applicable to both Cartesian and cylindrical coordinate systems. From (6), the magnetic field $\mathbf{H}$ is found to be [5]

$$\mathbf{H} = \frac{j}{\omega_0 \mu_0}\left(\nabla_T + \mathbf{a}_z \frac{\partial}{\partial z}\right) \times (\mathbf{E}_T + \mathbf{E}_z) = \frac{j}{\omega_0 \mu_0}\left(\nabla_T \times \mathbf{E}_T + \mathbf{a}_z \frac{\partial}{\partial z} \times \mathbf{E}_T + \nabla_T \times \mathbf{E}_z + \mathbf{a}_z \frac{\partial}{\partial z} \times \mathbf{E}_z\right) \tag{8}$$

Because of the requirement for transverse field components by the Poynting vector in (3), only those contributing to the transverse magnetic field in (8) are required. The first term is undesirable, since it only yields longitudinal field components. The last term vanishes. Then the required transverse magnetic field has the form

$$\mathbf{H}_T = \frac{j}{\omega_0 \mu_0}\left[\mathbf{a}_z \frac{\partial}{\partial z} \times (\mathbf{E}_{T_1} + \mathbf{E}_{T_2}) + \nabla_T \times \mathbf{E}_z\right] \tag{9}$$

Simplifying in column vector formalism

$$\mathbf{H}_T = \frac{j}{\omega_0 \mu_0}\left[|\mathbf{a}_z\rangle \frac{\partial}{\partial z} \times \begin{bmatrix} E_{T_1}|\mathbf{a}_{T_1}\rangle \\ E_{T_2}|\mathbf{a}_{T_2}\rangle \end{bmatrix} + \begin{bmatrix} |\mathbf{a}_{T_1}\rangle h_1^{-1} \partial/\partial r_{T_1} \\ |\mathbf{a}_{T_2}\rangle h_2^{-2} \partial/\partial r_{T_2} \end{bmatrix} \times E_z|\mathbf{a}_z\rangle \right] = \begin{bmatrix} H_{T_2}|\mathbf{a}_{T_2}\rangle \\ -H_{T_1}|\mathbf{a}_{T_1}\rangle \end{bmatrix} \tag{10}$$

$$\Rightarrow |H\rangle = \begin{bmatrix} |H_{T_2}\rangle \\ -|H_{T_1}\rangle \end{bmatrix} \tag{11}$$

Although correct, it can be seen that the Jones vector $|H\rangle$ is no longer in the canonical form previously described by (5). To see how it might be related to its canonical version, a simple modification is made to (11) that utilizes a specific Pauli matrix,

$$|H\rangle = \begin{bmatrix} |H_{T_2}\rangle \\ -|H_{T_1}\rangle \end{bmatrix} = \begin{pmatrix} 0 & 1 \\ -1 & 0 \end{pmatrix} \begin{bmatrix} |H_{T_1}\rangle \\ |H_{T_2}\rangle \end{bmatrix} = j\sigma_3 \begin{bmatrix} |H_{T_1}\rangle \\ |H_{T_2}\rangle \end{bmatrix} = j\sigma_3 |H_T\rangle \qquad (12)$$

where $j\sigma_3$ is a real matrix. The Pauli matrices are defined as the set [3]

$$\sigma_1 = \begin{pmatrix} 1 & 0 \\ 0 & -1 \end{pmatrix}, \quad \sigma_2 = \begin{pmatrix} 0 & 1 \\ 1 & 0 \end{pmatrix}, \quad \sigma_3 = \begin{pmatrix} 0 & -j \\ j & 0 \end{pmatrix}; \quad \sigma_0 = \begin{pmatrix} 1 & 0 \\ 0 & 1 \end{pmatrix} \qquad (13)$$

Pauli matrices are Hermitian matrices. Any arbitrary 2 x 2 complex matrix M may be expressed as a linear combination of the members of this set [3],

$$M = \tfrac{1}{2}a_0\sigma_0 + \tfrac{1}{2}a_1\sigma_1 + \tfrac{1}{2}a_2\sigma_2 + \tfrac{1}{2}a_3\sigma_3 \qquad (14)$$

for which the coefficients are derived using the trace (Tr) operator as follows

$$a_n = \mathrm{Tr}(\sigma_n M); \quad n \in \{0,1,2,3\} \qquad (15)$$

An identity relevant to this report is the following [3]

$$\sigma_m \sigma_m^\dagger = \sigma_m^\dagger \sigma_m = \sigma_0; \quad m \in \{1,2,3\} \qquad (16)$$

due to the fact that the Pauli matrices are all unitary.

The complex Poynting vector is found using one of the following expressions:

$$\mathbf{E}^* \times \mathbf{H} \cdot \mathbf{a}_z = E_{T_1}^* H_{T_2} - E_{T_2}^* H_{T_1} = \langle E_{T_1} | H_{T_2} \rangle - \langle E_{T_2} | H_{T_1} \rangle = \langle E_T | j\sigma_3 | H_T \rangle \qquad (17)$$

so that the power integral (3) becomes

$$P = \frac{1}{2}\int \langle E|H\rangle dS = \frac{1}{2}\int \langle E_T | j\sigma_3 | H_T \rangle dS \neq \frac{1}{2}\int \langle E_T | H_T \rangle dS \qquad (18)$$

Thus, power based on the cross-product is generally different from that found using the inner product, if the canonical form (5) is assumed.

The total field propagating in an optical fiber is actually the sum of the fields in the core and in the cladding,

$$|U\rangle = |U_J\rangle f_J + |U_K\rangle f_K : \quad |U\rangle \in \{|E\rangle, |H\rangle\} \qquad (19)$$

In (19), the subscript *J* is used to connote a Bessel function of the first kind (BFFK), which is used to describe the field in the core. The second subscript *K* corresponds to the modified Bessel function of the second kind (mBFSK), which describes the field in the geometrically annular cladding [2]. The geometrical basis functions used in (19) are

$$f_J = \text{rect}\left(\frac{\rho - a/2}{a}\right); \quad f_K = \text{rect}\left(\frac{\rho - (a+b)/2}{b - a}\right) \tag{20}$$

which respectively describe the geometries of the core and the cladding, and use the rectangle-function [6]. The 1st function describes a core of radius *a* and area $S_J$. The 2nd function represents the annular cladding of radial thickness (*b-a*), and area $S_K$. The orthogonality of the 2 basis functions can be succinctly stated as

$$f_\alpha f_\beta = \delta_{\alpha\beta} f_\alpha \tag{21}$$

The 2 functions are also linearly independent since they are orthogonal, which may be verified by their 2x2 Wronskian $W(f_J, f_K)$ over the circular region [0, *b*] [7]. Then using (19) in (18)

$$P = \frac{1}{2}\int \langle E | j\sigma_3 | H \rangle dS \tag{22}$$

simplifies to

$$P = \frac{1}{2}\int \left[ f_J \langle E_J | j\sigma_3 | H_J \rangle f_J + f_J \langle E_J | j\sigma_3 | H_K \rangle f_K + f_K \langle E_K | j\sigma_3 | H_J \rangle f_J + f_K \langle E_K | j\sigma_3 | H_K \rangle f_K \right] dS \tag{23}$$

where a typical term is found to be with the help of (21)

$$\int f_\alpha \langle E_\alpha | j\sigma_3 | H_\beta \rangle f_\beta dS = \delta_{\alpha\beta} \int_{S_\alpha} \langle E_\alpha | j\sigma_3 | H_\beta \rangle dS \tag{24}$$

which simplifies (23) to

$$P = \frac{1}{2}\int_{S_J} \langle E_J | j\sigma_3 | H_J \rangle dS + \frac{1}{2}\int_{S_K} \langle E_K | j\sigma_3 | H_K \rangle dS \tag{25}$$

The 2 terms respectively represent the powers in the core and in the cladding.

In the following sections, expressions for the EM powers in the core and the cladding of a step-index fiber are developed from the expressions of the transverse EM field, which are derived from Maxwell's equations, after finding the longitudinal EM field [2].

## 2 Electromagnetic Power In the Core of the Fiber

The step-index optical fiber is comprised of a core of radius $a$, and an annular cladding of a radial thickness of $(b-a)$. The refractive index of the core is $n_1$. The transverse field components are found using Maxwell's equations, and the longitudinal field components, which are in turn found from the longitudinal Helmholtz wave equation, a Bessel differential equation [2]. The field components are derived in [1, 2], and will not be re-derived here. The longitudinal field components can be expressed as complex quantities

$$E_z = jA\, J_n\left(\frac{u}{a}\rho\right)e^{-j(n\varphi+\psi)} \tag{26}$$

$$H_z = \frac{A\beta s}{\omega_0\mu_0} J_n\left(\frac{u}{a}\rho\right)e^{-j(n\varphi+\psi)} \tag{27}$$

In the core, the transverse electric field of the hybrid mode is comprised of radial ($E_\rho$) and azimuthal ($E_\varphi$) components, each of which is given by a weighted superposition of Bessel functions of the first kind (BFFK), in a product with a trigonometric function of the azimuth $\varphi$, as follows [2]

$$|E_\rho\rangle = -jA\beta\xi\left[(1-s)J_{n-1}\left(\frac{u}{a}\rho\right)-(1+s)J_{n+1}\left(\frac{u}{a}\rho\right)\right]\cos(n\varphi+\psi) \tag{28}$$

$$|E_\varphi\rangle = jA\beta\xi\left[(1-s)J_{n-1}\left(\frac{u}{a}\rho\right)+(1+s)J_{n+1}\left(\frac{u}{a}\rho\right)\right]\sin(n\varphi+\psi) \tag{29}$$

It is observed that the radial component is a subtractive expression, while that for the azimuth, is an additive expression. These observations will be later used in the derivation of the power expression. In this form, a transverse field component is also observed to be a linear combination of *non-orthogonal* functions, due to the inner product for two BFFKs of dissimilar orders being [8]

$$\int_0^a J_{n+1}(u\rho/a)J_{n-1}(u\rho/a)\rho\, d\rho = \frac{(2n+2)^{-1}\left(au^n/2^n\right)^2}{(n+1)!(n-1)!} {}_3F_4\left(\begin{array}{c}\frac{2n+1}{2},n+1,n+1;\\ n+2,n+2,n,2n+1;\end{array} -u^2\right) \tag{30}$$

which involves the generalized hypergeometric function ${}_3F_4$. The constant $A$ is an amplitude coefficient, $\beta$ is the propagation constant or the eigenvalue, and

$$u = a\left(k_0^2 n_1^2 - \beta^2\right)^{1/2} \tag{31}$$

$$w = a\left(\beta^2 - k_0^2 n_2^2\right)^{1/2} \tag{32}$$

$$v^2 = u^2 + w^2 \tag{33}$$

$$s = \frac{n\left(\frac{1}{u^2} + \frac{1}{w^2}\right)}{\left[\frac{J'_n(u)}{uJ_n(u)} + \frac{K'_n(w)}{wK_n(w)}\right]} = \frac{n\, v^2 J_n(u) K_n(w)}{uw^2 J'_n(u) K_n(w) + u^2 w K'_n(w) J_n(u)} \quad (34)$$

$$\xi = \frac{a}{2u} \quad (35)$$

The parameter $s$ occurs frequently throughout this report, and is seen to carry both the BFFK, the mBFSK, and their first-order derivatives. It will later be recognized to be an expression of the dispersion relation for the hybrid modes. Eqs. (31) and (32) are known as the normalized transverse wave numbers as they are unit-less. The parameter $v$ (33) is termed the *normalized frequency*. Eq. (33) is identical to that found for a 3-layer slab waveguide [2]. Any hybrid mode is capable of 2 polarization states, depending on the value of the angle

$$\psi = (m-1)\pi/2; \quad m \in \{1,2\} \quad (36)$$

where the x-polarization is represented by $m=1$, whereas the y-polarization, by $m=2$. The field components may be recast as quasi-vectors as follows,

$$|E_\varphi\rangle = -A\beta\xi \begin{bmatrix} (1-s) J_{n-1}(u\rho/a) \\ (1+s) J_{n+1}(u\rho/a) \end{bmatrix} e^{-j(n\varphi+\psi)} = -A\beta\xi |G_J(\rho,\varphi)\rangle; \quad (37)$$

$$|E_\rho\rangle = -jA\beta\xi \begin{bmatrix} (1-s) J_{n-1}(u\rho/a) \\ -(1+s) J_{n+1}(u\rho/a) \end{bmatrix} e^{-j(n\varphi+\psi)} = -j\sigma_1 A\beta\xi |G_J(\rho,\varphi)\rangle = j\sigma_1 |E_\varphi\rangle \quad (38)$$

where $\sigma_1$ is a Pauli matrix from the set (13), and

$$|G_J(\rho,\varphi)\rangle = \begin{bmatrix} (1-s) J_{n-1}(u\rho/a) \\ (1+s) J_{n+1}(u\rho/a) \end{bmatrix} e^{-j(n\varphi+\psi)} \quad (39)$$

and as a column vector, the transverse electric field may be expressed as

$$|E_T\rangle = \begin{bmatrix} |E_\rho\rangle \\ |E_\varphi\rangle \end{bmatrix} = \begin{bmatrix} j\sigma_1 \\ \sigma_0 \end{bmatrix} |E_\varphi\rangle = -\begin{bmatrix} j\sigma_1 \\ \sigma_0 \end{bmatrix} A\beta\xi |G_J\rangle \quad (40)$$

for which the dependence $(\rho,\varphi)$ has been suppressed for compactness.

The transverse magnetic field is given by [2],

$$|H_\rho\rangle = -jA\omega_0\varepsilon_1 \left(\frac{a}{2u}\right)\left[(1-s_1) J_{n-1}\left(\frac{u}{a}\rho\right) + (1+s_1) J_{n+1}\left(\frac{u}{a}\rho\right)\right] \sin(n\varphi+\psi) \quad (41)$$

$$\left|\mathrm{H}_{\varphi}\right\rangle = -jA\omega_0\varepsilon_1\left(\frac{a}{2u}\right)\left[(1-s_1)J_{n-1}\left(\frac{u}{a}\rho\right)-(1+s_1)J_{n+1}\left(\frac{u}{a}\rho\right)\right]\cos(n\varphi+\psi) \quad (42)$$

where $\varepsilon_1 = \varepsilon_0 n_1^2$, and (41, 42) are re-expressed as:

$$\left|\mathrm{H}_{\rho}\right\rangle = \varepsilon_1\omega_0 A\xi \begin{bmatrix}(1-s_1)J_{n-1}(u\rho/a)\\(1+s_1)J_{n+1}(u\rho/a)\end{bmatrix} e^{-j(n\varphi+\psi)} = \varepsilon_1\omega A\xi\left|G_J(\rho,\varphi)\right\rangle; \quad (43)$$

$$\left|\mathrm{H}_{\varphi}\right\rangle = -j\varepsilon_1\omega_0 A\xi \begin{bmatrix}(1-s_1)J_{n-1}(u\rho/a)\\-(1+s_1)J_{n+1}(u\rho/a)\end{bmatrix} e^{-j(n\varphi+\psi)} = -j\sigma_1\varepsilon_1\omega A\xi\left|G_J(\rho,\varphi)\right\rangle = -j\sigma_1\left|\mathrm{H}_{\rho}\right\rangle \quad (44)$$

where $s_1 = (\beta/k_0 n_1)^2 s$, and as a column vector, the transverse magnetic field is

$$\left|\mathrm{H}_T\right\rangle = \begin{bmatrix}\left|\mathrm{H}_{\rho}\right\rangle\\\left|\mathrm{H}_{\varphi}\right\rangle\end{bmatrix} = \begin{bmatrix}\sigma_0\\-j\sigma_1\end{bmatrix}\left|\mathrm{H}_{\rho}\right\rangle = \begin{bmatrix}\sigma_0\\-j\sigma_1\end{bmatrix}\varepsilon_1\omega_0 A\xi\left|G_J\right\rangle \quad (45)$$

Before proceeding farther with the derivation of the fiber core power, a justification is now given for the technique pursued thus far. From the previous section, the expression for EM power (17) was found to be a subtraction of $\left\langle\mathrm{E}_{\varphi}|\mathrm{H}_{\rho}\right\rangle$ from $\left\langle\mathrm{E}_{\rho}|\mathrm{H}_{\varphi}\right\rangle$. However, a cursory examination of the transverse field components (28, 29) and (41, 42) reveals that $\left\langle\mathrm{E}_{\varphi}|\mathrm{H}_{\rho}\right\rangle$ is negative whereas $\left\langle\mathrm{E}_{\rho}|\mathrm{H}_{\varphi}\right\rangle$ is positive, which transforms the subtraction in (17) into an addition. Further, $\left\langle\mathrm{E}_{\rho}|\mathrm{H}_{\varphi}\right\rangle$ is found from the product of 2 subtractive expressions, while $\left\langle\mathrm{E}_{\varphi}|\mathrm{H}_{\rho}\right\rangle$ is found from the product of 2 additive expressions. Therefore cross-terms of the form of $J_{n-1}(u\rho/a)J_{n+1}(u\rho/a)$ all cancel out upon the subtraction of $\left\langle\mathrm{E}_{\varphi}|\mathrm{H}_{\rho}\right\rangle$ from $\left\langle\mathrm{E}_{\rho}|\mathrm{H}_{\varphi}\right\rangle$ for the EM power. This argument effectively transforms the Cartesian product, into an inner product[1]. Recognizing this argument at the outset significantly simplifies the subsequent analysis; otherwise, the algebra can be tedious. The use of complex exponentials (instead of trigonometric functions), and a familiarity with Pauli matrices, also expedites the solution.

Using (18), the power bra-ket for the core becomes

$$P_J = \frac{1}{2}\int\left\langle\mathrm{E}_{\varphi}\right|[-j\sigma_1 \quad \sigma_0]|j\sigma_3|[\sigma_0 \quad -j\sigma_1]^T\left|\mathrm{H}_{\rho}\right\rangle dS = -\frac{1}{2}\int\left\langle\mathrm{E}_{\varphi}\right|2\sigma_0\left|\mathrm{H}_{\rho}\right\rangle dS \quad (46)$$

which was resolved with the help of (16). The Pauli matrix $\sigma_3$ from (13) is being used as an operator here, while the set {$-j\sigma_0$, $\sigma_1$} which excludes $\sigma_3$, is used as a quasi-basis for the transverse electric field, so that the product

---

[1] The Cartesian product of 2 sets {$a$, $b$} and {$x$, $y$} yields {$ax$, $ay$, $bx$, $by$}; whereas the inner product yields just {$ax$, $by$}

$$j\sigma_3 \begin{bmatrix} \sigma_0 \\ -j\sigma_1 \end{bmatrix} = \begin{bmatrix} 0 & 1 \\ -1 & 0 \end{bmatrix} \begin{bmatrix} \sigma_0 \\ -j\sigma_1 \end{bmatrix} = \begin{bmatrix} -j\sigma_1 \\ -\sigma_0 \end{bmatrix} \quad (47)$$

The bra-ket (46) evaluates to

$$P_J = -\frac{1}{2}\int \langle E_\varphi | 2\sigma_0 | H_\rho \rangle dS = -\int_0^{2\pi}\int_0^a \langle G_J | (-A\beta\xi)^\dagger \sigma_0 (\varepsilon_1 \omega A\xi) | G_J \rangle \rho d\rho d\varphi \quad (48)$$

Carrying out the integral over the azimuth $\varphi$ using (39):

$$P_J = 2\pi\varepsilon_1\omega_0\beta\xi^2 |A|^2 \int_0^a (1-s)(1-s_1)J_{n-1}^2(u\rho/a) + (1+s)(1+s_1)J_{n+1}^2(u\rho/a)\rho d\rho \quad (49)$$

Combining the 2 terms in the integrand using a $q$-summation,

$$P_J = 2\pi\varepsilon_1\omega_0\beta\xi^2 |A|^2 \sum_{q=-1}^{1} Q(q,s_1) \int_0^a J_{n+q}^2(u\rho/a)\rho d\rho \quad (50)$$

where

$$Q(q,g) = q^2(1+qs)(1+qg) \quad (51)$$

The integral in (50) has the following result [2]:

$$\int_0^a J_m^2(u\rho/a)\rho d\rho = \frac{1}{2}\left[ \rho^2 J_m^2(u\rho/a) - \rho^2 J_{m-1}(u\rho/a)J_{m+1}(u\rho/a) \right]_{\rho=0}^{\rho=a} \quad (52)$$

The 2 terms within the 1st square bracket may be reduced to one using a $p$-summation,

$$\int_0^a J_m^2(u\rho/a)\rho d\rho = \frac{1}{2}\sum_{p=0}^{1} e^{ip\pi}\left[ \rho^2 J_{m-p}(u\rho/a)J_{m+p}(u\rho/a) \right]_{\rho=0}^{\rho=a} \quad (53)$$

and evaluating the bracket yields

$$\int_0^a J_m^2(u\rho/a)\rho d\rho = \frac{a^2}{2}\sum_{p=0}^{1} e^{ip\pi}J_{m-p}(u)J_{m+p}(u) \quad (54)$$

which is then substituted back into (50), with every $m$ replaced by a $n+q$:

$$P_J = \pi\varepsilon_1\omega_0\beta\xi^2 a^2 |A|^2 \sum_{p=0}^{1}\sum_{q=-1}^{1} Q(q,s_1)e^{ip\pi}J_{n-p+q}(u)J_{n+p+q}(u) \quad (55)$$

## 3 Electromagnetic Power In the Cladding of the Fiber

The step-index optical fiber is comprised of a core of radius $a$, concentric with an annular cladding of a radial thickness of $(b\text{-}a)$, and a refractive index of $n_2$. The longitudinal field components in the cladding will not be derived, but are presented here for completeness,

$$E_z = jA \frac{J_n(u)}{K_n(w)} K_n\left(\frac{w}{a}\rho\right) e^{-j(n\varphi+\psi)} \tag{56}$$

$$H_z = \frac{A\beta s}{\omega_0 \mu_0} \frac{J_n(u)}{K_n(w)} K_n\left(\frac{w}{a}\rho\right) e^{-j(n\varphi+\psi)} \tag{57}$$

In the cladding, a transverse electric field components is expressed as a weighted superposition of the modified Bessel function of the second kind (mBFSK), in a product with a trigonometric function of the azimuth $\varphi$ [2]

$$|E_\rho\rangle = -jA\beta\left(\frac{aJ_n(u)}{2wK_n(w)}\right)\left[(1-s)K_{n-1}\left(\frac{w}{a}\rho\right) + (1+s)K_{n+1}\left(\frac{w}{a}\rho\right)\right]\cos(n\varphi+\psi) \tag{58}$$

$$|E_\varphi\rangle = jA\beta\left(\frac{aJ_n(u)}{2wK_n(w)}\right)\left[(1-s)K_{n-1}\left(\frac{w}{a}\rho\right) - (1+s)K_{n+1}\left(\frac{w}{a}\rho\right)\right]\sin(n\varphi+\psi) \tag{59}$$

which are expressible in column-vector form as

$$|E_\rho\rangle = -jA\beta R_n \begin{bmatrix}(1-s)K_{n-1}(w\rho/a) \\ (1+s)K_{n+1}(w\rho/a)\end{bmatrix} e^{-j(n\varphi+\psi)} = -jA\beta R_n |G_K(\rho,\varphi)\rangle; \tag{60}$$

$$|E_\varphi\rangle = -A\beta R_n \begin{bmatrix}(1-s)K_{n-1}(w\rho/a) \\ -(1+s)K_{n+1}(w\rho/a)\end{bmatrix} e^{-j(n\varphi+\psi)} = -\sigma_1 A\beta R_n |G_K(\rho,\varphi)\rangle = -j\sigma_1 E_\rho \tag{61}$$

where

$$R_n = \frac{aJ_n(u)}{2wK_n(w)} \tag{62}$$

$$|G_K(\rho,\varphi)\rangle = \begin{bmatrix}(1-s)K_{n-1}(w\rho/a) \\ (1+s)K_{n+1}(w\rho/a)\end{bmatrix} e^{-j(n\varphi+\psi)} \tag{63}$$

The transverse electric field vector is finally

$$|E_T\rangle = \begin{bmatrix}E_\rho \\ E_\varphi\end{bmatrix} = \begin{bmatrix}\sigma_0 \\ -j\sigma_1\end{bmatrix} E_\rho = -\begin{bmatrix}\sigma_0 \\ -j\sigma_1\end{bmatrix} jA\beta R_n |G_K\rangle = \begin{bmatrix}-j\sigma_0 \\ -\sigma_1\end{bmatrix}|G_K\rangle \tag{64}$$

for which a dependence on $(\rho, \varphi)$ has been suppressed for compactness. The transverse magnetic field is comprised of the components [2]

$$\left|H_\rho\right\rangle = -j\omega_0\varepsilon_2 AR_n\left[(1-s_2)K_{n-1}\left(\frac{w}{a}\rho\right)-(1+s_2)K_{n+1}\left(\frac{w}{a}\rho\right)\right]\sin(n\varphi+\psi) \tag{65}$$

$$\left|H_\varphi\right\rangle = -j\omega_0\varepsilon_2 AR_n\left[(1-s_2)K_{n-1}\left(\frac{w}{a}\rho\right)+(1+s_2)K_{n+1}\left(\frac{w}{a}\rho\right)\right]\cos(n\varphi+\psi) \tag{66}$$

where $s_2 = (\beta/k_0 n_2)^2 s$. Eqs. (65, 66) can be re-expressed as

$$\left|H_\varphi\right\rangle = -j\omega_0\varepsilon_2 AR_n\begin{bmatrix}(1-s_2)K_{n-1}(w\rho/a)\\(1+s_2)K_{n+1}(w\rho/a)\end{bmatrix}e^{-j(n\varphi+\psi)} = -j\omega\varepsilon_2 AR_n\left|G_K(\rho,\varphi)\right\rangle; \tag{67}$$

$$\left|H_\rho\right\rangle = \omega_0\varepsilon_2 AR_n\begin{bmatrix}(1-s_2)K_{n-1}(w\rho/a)\\-(1+s_2)K_{n+1}(w\rho/a)\end{bmatrix}e^{-j(n\varphi+\psi)} = \sigma_1\omega\varepsilon_2 AR_n\left|G_K(\rho,\varphi)\right\rangle = j\sigma_1 H_\varphi \tag{68}$$

The transverse magnetic field vector is thus

$$\left|H_T\right\rangle = \begin{bmatrix}H_\rho\\H_\varphi\end{bmatrix} = \begin{bmatrix}j\sigma_1\\\sigma_0\end{bmatrix}H_\varphi = -\begin{bmatrix}j\sigma_1\\\sigma_0\end{bmatrix}j\omega_0\varepsilon_2 AR_n\left|G_K\right\rangle = \begin{bmatrix}\sigma_1\\-j\sigma_0\end{bmatrix}\omega_0\varepsilon_2 AR_n\left|G_K\right\rangle \tag{69}$$

The power in the cladding is then, using (18)

$$P_K = \frac{1}{2}\int\left\langle E_\rho\right|[\sigma_0\ j\sigma_1]|j\sigma_3|[j\sigma_1\ \sigma_0]^T\left|H_\varphi\right\rangle dS = \int_0^{2\pi}\int_a^b\left\langle G_K\right|(-jA\beta R_n)^\dagger(-j\omega\varepsilon_2 AR_n)\left|G_K\right\rangle\rho\,d\rho\,d\varphi \tag{70}$$

To find the power in the cladding, it is widely assumed that the cladding terminates radially at infinity, which may be an excellent approximation for many step-index fibers. However, no such assumption is made here, and the integral is carried out over the actual boundaries of the cladding. Using (63), and simplifying

$$\begin{aligned}P_K &= 2\pi\varepsilon_2\omega_0\beta|A|^2 R_n^2\int_a^b(1-s)(1-s_2)K_{n-1}^2(w\rho/a)+(1+s)(1+s_2)K_{n+1}^2(w\rho/a)\rho\,d\rho\\ &= 2\pi\varepsilon_2\omega_0\beta|A|^2 R_n^2\sum_{q=-1}^{1}\int_a^b Q(q,s_2)K_{n+q}^2(w\rho/a)\rho\,d\rho\end{aligned} \tag{71}$$

which made use of (51). The integral under the summation has the result

$$\int_a^b K_r^2(w\rho/a)\rho\,d\rho = \frac{1}{2}b^2\left[K_r^2(wb/a)-K_{r-1}(wb/a)K_{r+1}(wb/a)\right]-\frac{1}{2}a^2\left[K_r^2(w)-K_{r-1}(w)K_{r+1}(w)\right] \tag{72}$$

The simplification begins by reducing the 2 brackets to just one using an *m*-summation:

$$\frac{1}{2}\sum_{m=1}^{2}\left(a^{2-m}b^{m-1}e^{im\pi/2}\right)^{2}\left[K_{r}^{2}\left(wa^{2-m}b^{m-1}/a\right)-K_{r-1}\left(wa^{2-m}b^{m-1}/a\right)K_{r+1}\left(wa^{2-m}b^{m-1}/a\right)\right] \quad (73)$$

followed by reducing the resultant bracketed terms to a single term using another, $p$-summation, therefore yielding for the integral

$$\int_{a}^{b}K_{r}^{2}(w\rho/a)\rho d\rho=\frac{1}{2}\sum_{m=1}^{2}\sum_{p=0}^{1}\left(a^{2-m}b^{m-1}e^{im\pi/2}\right)^{2}e^{ip\pi}K_{r-p}\left(wa^{2-m}b^{m-1}/a\right)K_{r+p}\left(wa^{2-m}b^{m-1}/a\right) \quad (74)$$

After a manipulation,

$$\int_{a}^{b}K_{r}^{2}(w\rho/a)\rho d\rho=\frac{a^{2}}{2}\sum_{m=1}^{2}\sum_{p=0}^{1}\kappa_{m}^{2}e^{i(m+p)\pi}K_{r-p}(w\kappa_{m})K_{r+p}(w\kappa_{m}) \quad (75)$$

where

$$\kappa_{m}=(a/b)^{1-m} \quad (76)$$

Eq. (75) is then substituted back into (71) for the power in the cladding, with $r=n+q$

$$P_{K}=2\pi\varepsilon_{2}\omega_{0}\beta a^{2}|A|^{2}R_{n}^{2}\sum_{q=-1}^{1}\sum_{m=1}^{2}\sum_{p=0}^{1}\kappa_{m}^{2}e^{im\pi}e^{ip\pi}Q(q,s_{2})K_{n-p+q}(w\kappa_{m})K_{n+p+q}(w\kappa_{m}) \quad (77)$$

## 4 The Dispersion Relation

The EM field components found from Maxwell's equations, initially have unknown amplitude coefficients. Continuity is required for the tangential $\varphi$- and $z$-components of the EM field at the circular boundary between the core and the cladding, expressed as $\rho = a$. For the $\varphi$-components, this requirement equates the electric fields (29) and (59), and the magnetic fields (42) and (66). For the $z$-components, the electric fields (26) and (56) must be the same, as must the magnetic fields (27) and (57) [1, 2]. Forcing to zero the determinant of the 4 resultant equations yields the dispersion relation

$$\left[\frac{J_{n}'(u)}{uJ_{n}(u)}+\frac{K_{n}'(w)}{wK_{n}(w)}\right]\left[n_{1}^{2}\frac{J_{n}'(u)}{uJ_{n}(u)}+n_{2}^{2}\frac{K_{n}'(w)}{wK_{n}(w)}\right]=n^{2}\frac{\beta^{2}}{k_{0}^{2}}\left(\frac{1}{u^{2}}+\frac{1}{w^{2}}\right)^{2} \quad (78)$$

The variables $u$ (31) and $w$ (32) are also both functions of the propagation constant, or the eigenvalue $\beta$. The equation can generally be solved numerically for the eigenvalues $\beta$ of the transverse (TE and TM) modes, as well as the (EH and HE) hybrid modes. The eigenvalue is usually subscripted as $\beta_{nr}$. This is due to the fact that the BFFK of the $n$-th order $J_{n}(u)$ (which describes the radial behavior of the longitudinal EM field *in the core*) exhibits oscillatory behavior similar to that of a damped sinusoid that results in $r$ roots for a given order *n*. These roots are identified with the eigenvalue $β_{nr}$, which represents either the $TE_{nr}$, $TM_{nr}$, $EH_{nr}$, or $HE_{nr}$ mode. The eigenvalues for the meridional rays represented by the $TE_{nr}$ and $TM_{nr}$ modes, are found from (78) by setting *n*=0. For the skew-rays which are the hybrid $EH_{nr}$ and $HE_{nr}$ modes, *n* must be at least unity and (78) would have to be solved numerically for the eigenvalues.

Using (31) and (32), the following relation can be found:

$$\frac{\beta^2}{k_0^2}\left(\frac{1}{u^2}+\frac{1}{w^2}\right)=\frac{n_1^2}{u^2}+\frac{n_2^2}{u^2} \tag{79}$$

Upon substitution of the above equation into (78),

$$\left[\frac{J'_n(u)}{uJ_n(u)}+\frac{K'_n(w)}{wK_n(w)}\right]\left[\left(\frac{n_1}{n_2}\right)^2\frac{J'_n(u)}{uJ'_n(u)}+\frac{K'_n(w)}{wK_n(w)}\right]=n^2\left(\frac{1}{u^2}+\frac{1}{w^2}\right)\left(\left(\frac{n_1}{n_2}\right)^2\frac{1}{u^2}+\frac{1}{w^2}\right) \tag{80}$$

The equation is still dependent on the eignenvalue *β* due to the presence of the normalized transverse wave numbers *u* (31) and *w* (32). No simplification is attained without further approximations; although it is observable that the 2 terms on each side of (80), are identical within a factor of $(n_1/n_2)^2$. A practical approximation is that of the *weakly*-guided fiber [9], which requires that the ratio of the core and cladding refractive indices, to be close to unity

$$\frac{n_1}{n_2}\approx 1 \tag{81}$$

For practical optical communication applications, the refractive index difference is approximately $\Delta n/n_1$, and is typically < 1% [2]. When (81) is applied to the general dispersion relation (80), it yields a dispersion relation specialized to the hybrid modes:

$$\left[\frac{J'_n(u)}{uJ_n(u)}+\frac{K'_n(w)}{wK_n(w)}\right]^2\approx n^2\left(\frac{1}{u^2}+\frac{1}{w^2}\right)^2 ; \ n\geq 1 \tag{82}$$

which results in 2 possible dispersion relations, as follows,

$$\frac{J'_n(u)}{uJ_n(u)}+\frac{K'_n(w)}{wK_n(w)}\approx \pm n\left(\frac{1}{u^2}+\frac{1}{w^2}\right); \ n\geq 1 \tag{83}$$

However, (82) is also recognized as the square of the unit-less parameter *s* from (34),

$$s^2\approx 1 \tag{84}$$

Thus *s* is effectively the dispersion relation for the hybrid modes of an optical fiber. The positive branch of (83), corresponding to $s \approx +1$, is assigned to the EH-modes. The negative branch of (83), equivalent to $s \approx -1$, is for the HE-modes. This has implications for other *s*-dependent parameters $s_1$ and $s_2$, collectively expressed by

$$s_p = \frac{\beta^2}{k_0^2 n_p^2} s = \frac{n_{\text{eff}}^2}{n_p^2} s; \quad p \in \{1,2\} \tag{85}$$

where $n_{\text{eff}}$ is the effective index of a propagating mode, and is obtained from its eigenvalue $\beta$. Since $n_{\text{eff}} \in [n_1, n_2]$ for a propagating mode, and in light of the weakly-guided fiber approximation (81),

$$s_p \approx s; \quad p \in \{1,2\} \tag{86}$$

Finally, it is concluded that in the weakly-guided-fiber approximation, for the HE-modes,

$$s_p \approx s \approx -1; \quad p \in \{1,2\} \tag{87}$$

and for the EH-modes

$$s_p \approx s \approx +1; \quad p \in \{1,2\} \tag{88}$$

The weakly-guided-fiber approximation permits the specialization of the developed power expressions to the HE- and EH-modes.

## 5 Total Electromagnetic Power In the Fiber

The total hybrid mode power flow in an optical fiber of core radius *a* and an annular cladding thickness of (*b-a*), based on (25), is given by the sum of the power in the core (55) and the power in the cladding (77),

$$P = P_J + P_K = \varepsilon_0 \omega_0 \beta P_0 S_J \left[ \begin{array}{c} n_1^2 \xi^2 \sum_{p=0}^{1} \sum_{q=-1}^{1} e^{ip\pi} Q(q,s_1) J_{n-p+q}(u) J_{n+p+q}(u) \\ + n_2^2 R_n^2 \sum_{m=1}^{2} \sum_{p=0}^{1} \sum_{q=-1}^{1} \kappa_m^2 e^{i(m+p)\pi} Q(q,s_2) K_{n-p+q}(w\kappa_m) K_{n+p+q}(w\kappa_m) \end{array} \right] \tag{89}$$

where $S_J$ is the cross-sectional area of the core. Eq. (89) yields a total of 10 terms. The terms for the core and cladding are easily identified by their pre-multiplicative refractive index factors, since $n_1$ was assigned to the core index, whereas $n_2$ was assigned to the *cladding* index. This is the most compact and general expression for hybrid modes, regardless of type, and prior to any approximations. The propagation constant $\beta$ is found from the dispersion relation (78), after which it is substituted into the parameters below (except (96)), before the total power (89) is obtained,

$$Q(q,g) = q^2(1+qs)(1+qg); \quad q \in \{-1,0,1\}, \ g \in \{s_1, s_2\} \tag{90}$$

$$s_p = \frac{\beta^2}{k_0^2 n_p^2} s; \quad p \in \{1,2\} \tag{91}$$

$$u = a\left(k_0^2 n_1^2 - \beta^2\right)^{1/2} \tag{92}$$

$$w = a\left(\beta^2 - k_0^2 n_2^2\right)^{1/2} \tag{93}$$

$$v^2 = u^2 + w^2 \tag{94}$$

$$s = \frac{nv^2 J_n(u) K_n(w)}{uw^2 J_n'(u) K_n(w) + u^2 w K_n'(w) J_n(u)} \tag{95}$$

$$\kappa_m = (a/b)^{1-m} \tag{96}$$

$$R_n = \frac{aJ_n(u)}{2wK_n(w)} \tag{97}$$

$$\xi = \frac{a}{2u} \tag{98}$$

and the peak power is given by

$$P_0 = |A|^2 \tag{99}$$

which assumes a CW field. Having a measurement of the average power permits a solution to (99) using (89),

$$P_0 = \frac{P}{\varepsilon_0 \omega_0 \beta S_J} \left[ \begin{array}{c} n_1^2 \xi^2 \sum_{p=0}^{1} \sum_{q=-1}^{1} e^{ip\pi} Q(q, s_1) J_{n-p+q}(u) J_{n+p+q}(u) \\ + n_2^2 R_n^2 \sum_{m=1}^{2} \sum_{p=0}^{1} \sum_{q=-1}^{1} \kappa_m^2 e^{i(m+p)\pi} Q(q, s_2) K_{n-p+q}(w\kappa_m) K_{n+p+q}(w\kappa_m) \end{array} \right]^{-1} \tag{100}$$

Using the dispersion relation (78) for the eigenvalue along with (89-100) should enable the power computation of a hybrid mode.

At this juncture, the power expression (89) bears no resemblance to other, widely used expressions [2]. However, simpler versions of (89) can be attained after applying the weakly-guided-fiber approximation (81), repeated here as the constraint

$$\frac{n_1}{n_2} \approx 1 \tag{101}$$

## 5.1 Power of HE-Modes for a Weakly-Guided Optical Fiber

To specialize the power expression (89) to the HE-modes for a weakly-guided optical fiber, Eq. (87) is invoked for $s \approx -1$, which reduces the $Q$-parameter (90) to the following expression,

$$Q(q,g) = q^2(1-q)(1+qg); \quad q \in \{-1,0,1\}, \quad g \in \{s_1, s_2\} \tag{102}$$

which has implications for the power expression (89): It reduces each $q$-summation to a single term since the $Q$-parameter in this case is zero unless $q = -1$. Moreover, $g \approx -1$ due to (87), so that

$$Q(-1,-1) = 4. \tag{103}$$

Then to summarize, invoking the weakly-guided-fiber approximation (101) for HE modes has 2 consequences: it reduces in (89) each $q$-summation to a single term, and reduces the $Q$-parameter to a constant (103). Thus, for any HE-mode, (89) simplifies to

$$P_{HE} = 4\varepsilon_0 \omega_0 \beta P_0 S_J \left[ \begin{array}{c} n_1^2 \left(\dfrac{a}{2u}\right)^2 \left[J_{n-1}^2(u) - J_{n-2}(u)J_n(u)\right] \\ + n_2^2 \left(\dfrac{aJ_n(u)}{2wK_n(w)}\right)^2 \sum_{m=1}^{2} \sum_{p=0}^{1} \kappa_m^2 e^{i(m+p)\pi} K_{n-p-1}(w\kappa_m) K_{n+p-1}(w\kappa_m) \end{array} \right] \tag{104}$$

which made use of (97) and (98). The expression may be further simplified by factoring out the core index $n_1$, due to (101), and simplifying

$$P_{HE} \approx \varepsilon_1 \omega_0 \beta P_0 S_J \left(\dfrac{a}{u}\right)^2 \left[ \begin{array}{c} \left(J_{n-1}^2(u) - J_{n-2}(u)J_n(u)\right) \\ + \left(\dfrac{uJ_n(u)}{wK_n(w)}\right)^2 \sum_{m=1}^{2} \sum_{p=0}^{1} \kappa_m^2 e^{i(m+p)\pi} K_{n-p-1}(w\kappa_m) K_{n+p-1}(w\kappa_m) \end{array} \right] \tag{105}$$

In order to simplify further, well-known recurrence relations for Bessel functions of the first kind (BFFK) and for modified Bessel functions of the second kind (mBFSK) are used, with $p$ being either 0 or 1:

$$\left(\dfrac{n}{u}\right)^p \dfrac{d^{1-p} J_n(u)}{du^{1-p}} = \dfrac{1}{2}\left(J_{n-1}(u) - e^{ip\pi} J_{n+1}(u)\right) \tag{106}$$

$$\left(\dfrac{n}{w}\right)^p \dfrac{d^{1-p} K_n(w)}{dw^{1-p}} = \dfrac{1}{2}\left(K_{n-1}(w) + e^{ip\pi} K_{n+1}(w)\right) \tag{107}$$

as well as the dispersion relation for the HE-modes [2],

$$\frac{J_{n-1}(u)}{J_n(u)} = +\frac{u}{w}\frac{K_{n-1}(w)}{K_n(w)} \tag{108}$$

which is obtained from the *negative* branch of the general dispersion relation for hybrid modes (83) - after using (106, 107) with *p=0* to eliminate the derivatives in (83), subsequently followed by applying (106, 107) again but with *p=1*, to eliminate Bessel functions of order (*n+1*).

The 1st parenthesized term within the square bracket in (105) can be re-expressed as

$$J_{n-1}^2(u) - J_{n-2}(u)J_n(u) = J_{n-1}^2(u)\left(1 - \frac{J_{n-2}(u)}{J_{n-1}(u)}\frac{J_n(u)}{J_{n-1}(u)}\right) = J_{n-1}^2(u)\left(1 + \left(\frac{w}{u}\right)^2 \frac{K_{n-2}(w)K_n(w)}{K_{n-1}^2(w)}\right) \tag{109}$$

There are 2 BFFK-ratios parenthesized in the middle equation above. The 1st term is resolved, by using (106) with *p=1* and *n* replaced by *n-1*, and solving the resultant equation simultaneously with (108) to eliminate $J_n(u)$. The result is then further simplified by using (107) with *p=1* and *n* replaced by *n-1*, to eliminate a term not dependent on Bessel functions. The 2nd BFFK ratio parenthesized in the middle equation above can be turned into a ratio in terms of mBFSKs using the inverse of (108).

The multiplicative factor of the double-summation term in (105) can also be re-cast using the dispersion relation (108), to

$$\left(\frac{uJ_n(u)}{wK_n(w)}\right)^2 = \left(\frac{J_{n-1}(u)}{K_{n-1}(w)}\right)^2 \tag{110}$$

which was obtained from (108) by a simple algebraic manipulation. Substituting (109, 110) back into (105) and after factoring out a common $J_{n-1}^2(u)$-term, results in

$$P_{HE} \approx \varepsilon_1\omega_0\beta P_0 S_J \left(\frac{a}{u}\right)^2 J_{n-1}^2(u)\left[\begin{array}{c} 1 + \left(\dfrac{w}{u}\right)^2 \dfrac{K_{n-2}(w)K_n(w)}{K_{n-1}^2(w)} \\ + \sum_{m=1}^{2}\sum_{p=0}^{1}\kappa_m^2 e^{i(m+p)\pi}\dfrac{K_{n-p-1}(w\kappa_m)K_{n+p-1}(w\kappa_m)}{K_{n-1}^2(w)} \end{array}\right] \tag{111}$$

The expression may be simplified by expanding the summations, canceling common terms, factoring out a common term, then reconstituting the *p*-summation,

$$P_{HE} \approx \varepsilon_1\omega_0\beta P_0 S_J \left(\frac{va}{u^2}\right)^2 J_{n-1}^2(u)\frac{K_{n-2}(w)K_n(w)}{K_{n-1}^2(w)}\left[1 + \left(\frac{ub}{va}\right)^2 \sum_{p=0}^{1}\frac{e^{ip\pi}K_{n-p-1}(bw/a)K_{n+p-1}(bw/a)}{K_{n-2}(w)K_n(w)}\right] \tag{112}$$

which also made use of (96). For ITU-T G.652 single-mode fiber, such as Corning SMF28e+®[10], the core and cladding diameters respectively are $2a \approx 8$ μm and $2b \approx 125$ μm, which yields a $b/a \approx 16$, and renders the 2nd bracketed term relatively negligible. The resultant expression is then in agreement with that found in [2]. For ITU G.651 multi-mode sensor fiber such as Corning ClearCurve®[11], the core and cladding diameters respectively are $2a \approx 50$ μm and $2b \approx 125$ μm, which yields a $b/a \approx 2.5$, and the 2nd bracketed term may not be so negligible for some higher-order modes.

An important special case of (112) is the $HE_{11}$-mode, which is the only relevant mode for single-mode fibers, such as the Corning SMF28e+®, widely used in optical communication. Setting $n=1$ in the above expression, and making use of the following identity for the corresponding terms in (112)

$$K_p(x) = K_{-p}(x) = \frac{\pi}{2} \lim_{\nu \to p} \frac{I_{-\nu}(x) - I_\nu(x)}{\sin(\nu \pi)} \tag{113}$$

where $\nu$ (not to be confused with the normalized frequency $v$) denotes the fractional order of the modified Bessel function of the first kind $I$, reduces (112) to the following simple expression

$$P_{HE_{11}} \approx \varepsilon_1 \omega_0 \beta_{11} P_0 S_J \left(\frac{va}{u^2}\right)^2 J_0^2(u) \frac{K_1^2(w)}{K_0^2(w)} \left[1 + \left(\frac{ub}{va}\right)^2 \frac{K_0^2(bw/a) - K_1^2(bw/a)}{K_1^2(w)}\right] \approx \varepsilon_1 \omega_0 \beta_{11} P_0 S_J \left(\frac{va}{u^2}\right)^2 J_0^2(u) \frac{K_1^2(w)}{K_0^2(w)} \tag{114}$$

which agrees with the result found in [2].

**5.2 Power of EH-Modes for a Weakly-Guided Optical Fiber**
To specialize the power expression (89) to the EH-modes for a weakly-guided fiber, Eq. (88) is applied for $s \approx +1$, simplifying the $Q$-parameter (90) to

$$Q(q,g) = q^2(1+q)(1+qg); \quad q \in \{-1,0,1\}, \quad g \in \{s_1, s_2\} \tag{115}$$

which reduces each $q$-summation in (89) to a single term since (115) is zero unless $q = +1$. Moreover, $g \approx +1$ due to (91) regardless of what it is, so that

$$Q(+1,+1) = 4. \tag{116}$$

The expression for the EH-mode power is thus given by

$$P_{EH} = 4\varepsilon_0 \omega_0 \beta P_0 S_J \left[\begin{array}{c} n_1^2 \xi^2 \left(J_{n+1}^2(u) - J_n(u)J_{n+2}(u)\right) \\ +n_2^2 R_n^2 \sum_{m=1}^{2} \sum_{p=0}^{1} \kappa_m^2 e^{i(m+p)\pi} K_{n-p+1}(w\kappa_m) K_{n+p+1}(w\kappa_m) \end{array}\right] \tag{117}$$

which can be re-cast as, after factoring out the refractive index of the core due to (101), and making use of (97) and (98),

$$P_{EH} \approx \varepsilon_1 \omega_0 \beta P_0 S_J \left(\frac{a}{u}\right)^2 \left[ \begin{array}{c} \left(J_{n+1}^2(u) - J_n(u) J_{n+2}(u)\right) \\ + \left(\frac{u J_n(u)}{w K_n(w)}\right)^2 \sum_{m=1}^{2} \sum_{p=0}^{1} \kappa_m^2 e^{i(m+p)\pi} K_{n-p+1}(w\kappa_m) K_{n+p+1}(w\kappa_m) \end{array} \right] \quad (118)$$

The dispersion relation for the EH-modes,

$$\frac{J_{n+1}(u)}{J_n(u)} = -\frac{u}{w} \frac{K_{n+1}(w)}{K_n(w)} \quad (119)$$

is obtained from the *positive* branch of the general dispersion relation for hybrid modes (83) - after using (106, 107) with $p=0$ to eliminate the derivatives in (83), subsequently followed by applying (106, 107) again but with $p=1$, to eliminate Bessel functions of order ($n$-1).

The 1st parenthesized term within the square bracket in (118) can be re-expressed as

$$J_{n+1}^2(u) - J_n(u) J_{n+2}(u) = J_{n+1}^2(u) \left(1 - \frac{J_{n+2}(u)}{J_{n+1}(u)} \frac{J_n(u)}{J_{n+1}(u)}\right) = J_{n+1}^2(u) \left(1 + \left(\frac{w}{u}\right)^2 \frac{K_n(w) K_{n+2}(w)}{K_{n+1}^2(w)}\right) \quad (120)$$

There are 2 BFFK-ratios parenthesized in the middle equation above. The 1st term is resolved, by using (106) with $p=1$ and $n$ replaced by $n+1$ and solving the resultant equation simultaneously with (119) to eliminate $J_n(u)$. The result is further simplified by using (107) with $p=1$ and $n$ replaced by $n+1$, to eliminate a term not dependent on Bessel functions. The 2nd BFFK ratio parenthesized in the middle equation above was turned into another ratio but in terms of mBFSKs using the inverse of (119).

The multiplicative factor of the double-summation term in (118) can also be re-cast using the dispersion relation (119), to

$$\left(\frac{u J_n(u)}{w K_n(w)}\right)^2 = \left(\frac{J_{n+1}(u)}{K_{n+1}(w)}\right)^2 \quad (121)$$

which was obtained from (119) by a simple algebraic manipulation. Substituting (120, 121) into (118) and after factoring out a $J_{n+1}^2(u)$-term common to both (120, 121),

$$P_{EH} \approx 4 \varepsilon_1 \omega_0 \beta P_0 S_J J_{n+1}^2(u) \left[ \begin{array}{c} 1 + \left(\frac{w}{u}\right)^2 \frac{K_n(w) K_{n+2}(w)}{K_{n+1}^2(w)} \\ + \sum_{m=1}^{2} \sum_{p=0}^{1} \kappa_m^2 e^{i(m+p)\pi} \frac{K_{n-p+1}(w\kappa_m) K_{n+p+1}(w\kappa_m)}{K_{n+1}^2(w\kappa_m)} \end{array} \right] \quad (122)$$

which is identical to that for the HE-mode (111), but with $n$ replaced by $n+2$. It simplifies to the following expression:

$$P_{EH} \approx \varepsilon_1 \omega_0 \beta P_0 S_J \left(\frac{va}{u^2}\right)^2 J_{n+1}^2(u) \frac{K_n(w) K_{n+2}(w)}{K_{n+1}^2(w)} \left[1 + \left(\frac{ub}{va}\right)^2 \sum_{p=0}^{1} \frac{e^{ip\pi} K_{n-p+1}(bw/a) K_{n+p+1}(bw/a)}{K_n(w) K_{n+2}(w)}\right]$$

(123)

In a *multi-mode* fiber, the first EH-mode to propagate is the $EH_{11}$-mode, for which

$$P_{EH} \approx \varepsilon_1 \omega_0 \beta P_0 S_J \left(\frac{va}{u^2}\right)^2 J_2^2(u) \frac{K_1(w) K_3(w)}{K_2^2(w)} \left[1 + \left(\frac{ub}{va}\right)^2 \frac{K_2^2(bw/a) - K_1(bw/a) K_3(bw/a)}{K_1(w) K_3(w)}\right] \quad (124)$$

However, the $EH_{11}$-mode, unlike the $HE_{11}$-mode, is not even among the first 3 modes to propagate in a multi-mode fiber [2].

## 6 Summary and Conclusions

An exact, compact expression for the power of a hybrid mode of propagation constant $\beta$, for a step-index fiber of core radius $a$, and a cladding radial thickness of $(b-a)$, is found to be

$$P = \varepsilon_0 \omega_0 \beta P_0 S_J \left[ \begin{array}{c} n_1^2 \xi^2 \sum_{p=0}^{1} \sum_{q=-1}^{1} e^{ip\pi} Q(q, s_1) J_{n-p+q}(u) J_{n+p+q}(u) \\ + n_2^2 R_n^2 \sum_{m=1}^{2} \sum_{p=0}^{1} \sum_{q=-1}^{1} \kappa_m^2 e^{i(m+p)\pi} Q(q, s_2) K_{n-p+q}(w\kappa_m) K_{n+p+q}(w\kappa_m) \end{array} \right] \quad (125)$$

where $S_J$ is the cross-sectional area of the core. The expression (125) is seen to be the sum of the powers in the core (identified by its refractive index $n_1$), and that in the cladding (identified by its refractive index $n_2$). The expression is found from the transverse EM-field, after the derivation of the longitudinal EM-field using the longitudinal Hemlholtz wave equation [2]. The various parameters are

$$Q(q, g) = q^2 (1 + qs)(1 + qg); \quad q \in \{-1, 0, 1\}, \quad g \in \{s_1, s_2\} \quad (126)$$

$$s_p = \frac{\beta^2}{k_0^2 n_p^2} s; \quad p \in \{1, 2\} \quad (127)$$

$$s = \frac{n\left(\frac{1}{u^2} + \frac{1}{w^2}\right)}{\left[\frac{J_n'(u)}{u J_n(u)} + \frac{K_n'(w)}{w K_n(w)}\right]} = \frac{n v^2 J_n(u) K_n(w)}{uw^2 J_n'(u) K_n(w) + u^2 w K_n'(w) J_n(u)} \quad (128)$$

$$u = a\left(k_0^2 n_1^2 - \beta^2\right)^{1/2} \quad (129)$$

$$w = a\left(\beta^2 - k_0^2 n_2^2\right)^{1/2} \tag{130}$$

$$v^2 = u^2 + w^2 \tag{131}$$

$$R_n = \frac{aJ_n(u)}{2wK_n(w)} \tag{132}$$

$$\xi = \frac{a}{2u} \tag{133}$$

$$\kappa_m = (a/b)^{1-m} \tag{134}$$

which, with the exception of the last equation, are all dependent on the propagation constant or eigenvalue $\beta$, numerically determined from the dispersion relation [2]

$$\left[\frac{J_n'(u)}{uJ_n(u)} + \frac{K_n'(w)}{wK_n(w)}\right]\left[\left(\frac{n_1}{n_2}\right)^2 \frac{J_n'(u)}{uJ_n'(u)} + \frac{K_n'(w)}{wK_n(w)}\right] = n^2\left(\frac{1}{u^2} + \frac{1}{w^2}\right)\left(\left(\frac{n_1}{n_2}\right)^2 \frac{1}{u^2} + \frac{1}{w^2}\right) \tag{135}$$

In the weakly-guided-fiber approximation [9],

$$\frac{n_1}{n_2} \approx 1 \tag{136}$$

the general dispersion relation (135) reduces to that for the hybrid modes:

$$s = \frac{n \cdot \left(\dfrac{1}{u^2} + \dfrac{1}{w^2}\right)}{\left[\dfrac{J_n'(u)}{uJ_n(u)} + \dfrac{K_n'(w)}{wK_n(w)}\right]} \approx \pm 1 \tag{137}$$

Eq.(125) yields the power for an EH-mode when

$$\begin{cases} q = +1; \\ \lim_{\substack{s \to +1 \\ g \to +1}} Q(q,g) = 4 \end{cases} \tag{138}$$

which simplifies (125) to

$$P_{EH} \approx \varepsilon_1 \omega_0 \beta P_0 S_J \left(\frac{va}{u^2}\right)^2 J_{n+1}^2(u) \frac{K_n(w)K_{n+2}(w)}{K_{n+1}^2(w)}\left[1 + \left(\frac{ub}{va}\right)^2 \sum_{p=0}^{1} \frac{e^{ip\pi}K_{n-p+1}(bw/a)K_{n+p+1}(bw/a)}{K_n(w)K_{n+2}(w)}\right] \tag{139}$$

The power for the HE-mode is obtained when

$$\begin{cases} q = -1; \\ \lim_{\substack{s \to -1 \\ g \to -1}} Q(q,g) = 4 \end{cases} \quad (140)$$

simplifying (125) to

$$P_{HE} \approx \varepsilon_1 \omega_0 \beta P_0 S_J \left(\frac{va}{u^2}\right)^2 J_{n-1}^2(u) \frac{K_{n-2}(w) K_n(w)}{K_{n-1}^2(w)} \left[1 + \left(\frac{ub}{va}\right)^2 \sum_{p=0}^{1} \frac{e^{ip\pi} K_{n-p-1}(bw/a) K_{n+p-1}(bw/a)}{K_{n-2}(w) K_n(w)}\right] \quad (141)$$

For commercial, weakly-guided single-mode fiber, the fundamental $HE_{11}$-mode is found to have a power expression given by

$$P_{HE_{11}} \approx \varepsilon_1 \omega_0 \beta_{11} P_0 S_J \left(\frac{va}{u^2}\right)^2 J_0^2(u) \frac{K_1^2(w)}{K_0^2(w)} \left[1 + \left(\frac{ub}{va}\right)^2 \frac{K_0^2(bw/a) - K_1^2(bw/a)}{K_1^2(w)}\right] \approx \varepsilon_1 \omega_0 \beta_{11} P_0 S_J \left(\frac{va}{u^2}\right)^2 J_0^2(u) \frac{K_1^2(w)}{K_0^2(w)} \quad (142)$$

## References

[1] K. Kawano and T. Kitoh, *Introduction to Optical Waveguide Analysis*, Chap. 2. New York: John Wiley & Sons, Inc., 2001
[2] K. Okamoto, *Fundamentals of Optical Waveguides*, 2nd Ed., Chaps. 1, 3. Massachusetts: Academic Press (Elsevier), 2006
[3] J. N. Damask, *Polarization Optics in Telecommunications*, Chap. 2. https://www.springeronline.com, 2005
[4] E. W. Weisstein, "Gradient." From *MathWorld*, A Wolfram Web Resource. https://mathworld.wolfram.com/Gradient.html
[5] W. P. Huang, and C. L. Xu, "Simulation of three-dimensional optical waveguides by a full-vector beam propagation method," *IEEE J. Quantum Electron.*, vol. 29, no. 10, Oct. 1993, pp. 2639-2649
[6] E. W. Weisstein, "Rectangle Function." From *MathWorld*, A Wolfram Web Resource. https://mathworld.wolfram.com/RectangleFunction.html
[7] E. W. Weisstein, "Linearly Dependent Functions." From *MathWorld*, A Wolfram Web Resource. https://mathworld.wolfram.com/LinearlyDependentFunctions.html
[8] https://www.wolframalpha.com/input/?i=besselJ%5Bnu%2Cx%5D*besselJ%5Bmu%2Cx%5D*x+
[9] D. Gloge, "Weakly guiding fibers," *Applied Optics*, vol. 10, no. 10, Oct. 1971, pp. 2252-2258
[10] Corning® SMF28e+® Product Information, July 2014: https://www.corning.com/media/worldwide/coc/documents/PI1463_07-14_English.pdf
[11] Corning® ClearCurve® Product Information, February 2013: https://www.corning.com/microsites/coc/oem/documents/specialty-fiber/Corning-Specialty-Fiber-Product-Information-Sheets-111913.pdf